\documentclass[preprint,12pt]{elsarticle}
\usepackage{amssymb}

\usepackage[ruled,noline]{algorithm2e}
\usepackage{extarrows}
\usepackage{amsmath}
\usepackage{mathrsfs}
\usepackage{bbm}
\usepackage{booktabs}
\usepackage{subfigure}
\makeatletter
\def\ps@pprintTitle{%
  \let\@oddhead\@empty
  \let\@evenhead\@empty
  \def\@oddfoot{\reset@font\hfil\thepage\hfil}
  \let\@evenfoot\@oddfoot
}
\makeatother

\makeatletter

\newcommand {\Rmnum} [1] {\expandafter \@slowromancap \romannumeral #1@}
\makeatother

\newtheorem{protocol}{\noindent\bf PROCOTOL}

\begin{document}

\begin{frontmatter}

\title{The Classification of Quantum Symmetric-Key Encryption Protocols}


\author[author1]{Chong Xiang}
\author[author2,author3]{Li Yang\corref{cor1}}
\author[author1]{Yong Peng}
\author[author1]{Dongqing Chen}

\cortext[cor1]{Corresponding author. E-mail: yangli@iie.ac.cn}
\address[author1]{China Information Technology Security Evaluation Center, Beijing 100085, China}
\address[author2]{State Key Laboratory of Information Security, Institute of Information Engineering, Chinese Academy of Sciences, Beijing 100093, China}
\address[author3]{Data Assurance and Communication Security Research Center,Chinese Academy of Sciences, Beijing 100093, China}
\begin{abstract}
The classification of quantum symmetric-key encryption protocol is presented.
According to five elements of a quantum symmetric-key encryption protocol: plaintext, ciphertext, key,
encryption algorithm and decryption algorithm, there are 32 different kinds of them. Among them, 5 kinds
of protocols have already been constructed and studied, and 21 kinds of them are proved to be impossible to
construct,
the last 6 kinds of them are not yet presented effectively. That means the research on quantum symmetric-key
encryption protocol only needs to consider with 5 kinds of them nowadays.

\end{abstract}


\end{frontmatter}



\section{Introduction}
In the study of quantum cryptography, it is generally agreed that the classical cryptography is a special case of the quantum cryptography.
When it comes to quantum encryption protocol, there are usually two viewpoints: the narrow one says that only the encryption algorithm involves quantum parts can be called quantum encryption protocol,
which is regarded as a counterpart to the classical encryption protocol; the generalized one says that the classical encryption protocol is a special case of quantum encryption protocol,
so the concept of quantum encryption protocol should be including the classical encryption protocol, and while it has nothing to do with a quantum part, it is called a classical one.
In this article, we take the second view to explain the quantum encryption protocol. If a quantum operation involves a classical target $x$,
it means we use the orthogonal basis to code quantum state $|x\rangle$ as the target.

Besides the classical encryption protocol, the most famous quantum encryption protocol is private quantum channel(PQC).
It was Boykin\cite{Boykin03} who firstly suggested the quantum one time pad, he encrypted $n$ qubits by $2n$ bits classical key with Pauli rotation operations.
Then he proved that $2n$ bits classical key was the necessary and sufficient condition of encrypting $n$ qubits with unconditional security. Ambainis\cite{Ambainis00} presented the definition of PQC,
and proved that two parties share $2n$ bits classical key is the necessary and sufficient condition of transferring $n$ qubits in public channel with unconditional security,
in this manner each qubit is encrypted to a ultimate mixed state. Based on this protocol, Ambainis\cite{Ambainis04} and Hardy\cite{hayden2004randomizing} separately presented
a quantum approximate encryption schemes with short keys. Bostrom suggested a protocol, which encrypts classical message with a entangled-state-based quantum algorithm\cite{bostrom2002deterministic}.
\section{Classification of the Quantum symmetric-key Encryption Protocol}
A quantum symmetric-key encryption protocol with five elements
 $(P,C,K,$
 $E,D):$ plaintext$(P)$, ciphertext$(C)$, key$(K)$, encryption algorithm$(E)$ and decryption algorithm$(D)$, can be divided into 32 classes according to
the property of each element. Among them, 5 kinds have already been constructed and studied, and 21 kinds are
proved to be impossible to be construct, the last 6 kinds are not yet presented effectively. That means the
research on quantum symmetric-key encryption protocol only needs to consider with 5 kinds of them nowadays. The specific classification is shown in Tab. 1.

\begin{table}[htbp]
  \centering
  \caption{32 kinds of quantum symmetric-key encryption protocols}
    \begin{tabular}{|c|ccccc|c|}
    \toprule
    {\bf Kind}      & ($P$     & $C$     & $K$     & $E$     & $D$)   & existence  \\
    \midrule
    1     & $\mathbb{C}$ & $\mathbb{C}$ & $\mathbb{C}$ & $\mathbb{C}$ & $\mathbb{C}$ & $\mathbb{E}$ \\
    2     & $\mathbb{C}$ & $\mathbb{C}$ & $\mathbb{C}$ & $\mathbb{C}$ & $\mathbb{Q}$ & $\mathbb{O}$ \\
    3     & $\mathbb{C}$ & $\mathbb{C}$ & $\mathbb{C}$ & $\mathbb{Q}$ & $\mathbb{C}$ & $\mathbb{O}$ \\
    4     & $\mathbb{C}$ & $\mathbb{C}$ & $\mathbb{C}$ & $\mathbb{Q}$ & $\mathbb{Q}$ & $\mathbb{O}$ \\
    5     & $\mathbb{C}$ & $\mathbb{C}$ & $\mathbb{Q}$ & $\mathbb{C}$ & $\mathbb{C}$ & $\mathbb{N}$ \\
    6     & $\mathbb{C}$ & $\mathbb{C}$ & $\mathbb{Q}$ & $\mathbb{C}$ & $\mathbb{Q}$ & $\mathbb{N}$ \\
    7     & $\mathbb{C}$ & $\mathbb{C}$ & $\mathbb{Q}$ & $\mathbb{Q}$ & $\mathbb{C}$ & $\mathbb{N}$ \\
    8     & $\mathbb{C}$ & $\mathbb{C}$ & $\mathbb{Q}$ & $\mathbb{Q}$ & $\mathbb{Q}$ & $\mathbb{O}$ \\
    9     & $\mathbb{C}$ & $\mathbb{Q}$ & $\mathbb{C}$ & $\mathbb{C}$ & $\mathbb{C}$ & $\mathbb{N}$ \\
    10    & $\mathbb{C}$ & $\mathbb{Q}$ & $\mathbb{C}$ & $\mathbb{C}$ & $\mathbb{Q}$ & $\mathbb{N}$ \\
    11    & $\mathbb{C}$ & $\mathbb{Q}$ & $\mathbb{C}$ & $\mathbb{Q}$ & $\mathbb{C}$ & $\mathbb{N}$ \\
    12    & $\mathbb{C}$ & $\mathbb{Q}$ & $\mathbb{C}$ & $\mathbb{Q}$ & $\mathbb{Q}$ & $\mathbb{E}$ \\
    13    & $\mathbb{C}$ & $\mathbb{Q}$ & $\mathbb{Q}$ & $\mathbb{C}$ & $\mathbb{C}$ & $\mathbb{N}$ \\
    14    & $\mathbb{C}$ & $\mathbb{Q}$ & $\mathbb{Q}$ & $\mathbb{C}$ & $\mathbb{Q}$ & $\mathbb{N}$ \\
    15    & $\mathbb{C}$ & $\mathbb{Q}$ & $\mathbb{Q}$ & $\mathbb{Q}$ & $\mathbb{C}$ & $\mathbb{N}$ \\
    16    & $\mathbb{C}$ & $\mathbb{Q}$ & $\mathbb{Q}$ & $\mathbb{Q}$ & $\mathbb{Q}$ & $\mathbb{E}$ \\
    17    & $\mathbb{Q}$ & $\mathbb{C}$ & $\mathbb{C}$ & $\mathbb{C}$ & $\mathbb{C}$ & $\mathbb{N}$ \\
    18    & $\mathbb{Q}$ & $\mathbb{C}$ & $\mathbb{C}$ & $\mathbb{C}$ & $\mathbb{Q}$ & $\mathbb{N}$ \\
    19    & $\mathbb{Q}$ & $\mathbb{C}$ & $\mathbb{C}$ & $\mathbb{Q}$ & $\mathbb{C}$ & $\mathbb{N}$ \\
    20    & $\mathbb{Q}$ & $\mathbb{C}$ & $\mathbb{C}$ & $\mathbb{Q}$ & $\mathbb{Q}$ & $\mathbb{O}$ \\
    21    & $\mathbb{Q}$ & $\mathbb{C}$ & $\mathbb{Q}$ & $\mathbb{C}$ & $\mathbb{C}$ & $\mathbb{N}$ \\
    22    & $\mathbb{Q}$ & $\mathbb{C}$ & $\mathbb{Q}$ & $\mathbb{C}$ & $\mathbb{Q}$ & $\mathbb{N}$ \\
    23    & $\mathbb{Q}$ & $\mathbb{C}$ & $\mathbb{Q}$ & $\mathbb{Q}$ & $\mathbb{C}$ & $\mathbb{N}$ \\
    24    & $\mathbb{Q}$ & $\mathbb{C}$ & $\mathbb{Q}$ & $\mathbb{Q}$ & $\mathbb{Q}$ & $\mathbb{O}$ \\
    25    & $\mathbb{Q}$ & $\mathbb{Q}$ & $\mathbb{C}$ & $\mathbb{C}$ & $\mathbb{C}$ & $\mathbb{N}$ \\
    26    & $\mathbb{Q}$ & $\mathbb{Q}$ & $\mathbb{C}$ & $\mathbb{C}$ & $\mathbb{Q}$ & $\mathbb{N}$ \\
    27    & $\mathbb{Q}$ & $\mathbb{Q}$ & $\mathbb{C}$ & $\mathbb{Q}$ & $\mathbb{C}$ & $\mathbb{N}$ \\
    28    & $\mathbb{Q}$ & $\mathbb{Q}$ & $\mathbb{C}$ & $\mathbb{Q}$ & $\mathbb{Q}$ & $\mathbb{E}$ \\
    29    & $\mathbb{Q}$ & $\mathbb{Q}$ & $\mathbb{Q}$ & $\mathbb{C}$ & $\mathbb{C}$ & $\mathbb{N}$ \\
    30    & $\mathbb{Q}$ & $\mathbb{Q}$ & $\mathbb{Q}$ & $\mathbb{C}$ & $\mathbb{Q}$ & $\mathbb{N}$ \\
    31    & $\mathbb{Q}$ & $\mathbb{Q}$ & $\mathbb{Q}$ & $\mathbb{Q}$ & $\mathbb{C}$ & $\mathbb{N}$ \\
    32    & $\mathbb{Q}$ & $\mathbb{Q}$ & $\mathbb{Q}$ & $\mathbb{Q}$ & $\mathbb{Q}$ & $\mathbb{E}$ \\
    \bottomrule
    \end{tabular}%
  \label{tab:classification}%
\end{table}%

In Tab. 1, $\mathbb{C}$ denotes the element belongs to classical space, $\mathbb{Q}$ denotes the element belongs to quantum space, $\mathbb{E}$ means the protocol is exist,
$\mathbb{N}$ means the protocol is not exist, $\mathbb{O}$ means whether this kind of protocol exists or not is still an open problem. According to this,
we classification the quantum symmetric-key encryption protocols into three types.
\section{Type $\mathbb{E}$}
In this section, five kinds of protocols of type $\mathbb{E}$ will be introduced through simple examples, the sequence numbers of those protocol
are Kind 1, 12, 16, 28, 32.
\subsection{{\bf Kind} 1}
Since each element of Kind 1 protocol belongs to classical space, this kind of quantum symmetric-key encryption protocol is actually the classical symmetric-key encryption.
There is no doubt that this kind is exist. DES\cite{biham1991des,Biham1992Differential}, AES\cite{daemen2002aes} and other common symmetric-key encryption protocols all belong to this kind.
\subsection{{\bf Kind} 12}
The Kind 12 protocol requests: $P,K\in\mathbb{C};~~C,E,D\in\mathbb{Q}$. This kind of protocol is relatively researched widely, we give one simple example for it as follow:
\begin{protocol}[$P,K\in\mathbb{C};~~C,E,D\in\mathbb{Q}$]

Let key be $k=(k_1,k_2)$, and classical plaintext be $x$.

{\bf Encryption: }\begin{enumerate}
                  \item Alice prepares the quantum state $|x\rangle_0$ according to $x$.
                  \item Alice performs quantum operations $Y^{k_2}H^{k_1}$ on $|x\rangle_0$. 

                      here $|\varphi_x\rangle=Y^{k_2}H^{k_1}|x\rangle_0=(-1)^{x\cdot k_2}|x\oplus k_2\rangle_{k_1}$。
                  \item Alice sends $|\varphi_x\rangle$ to Bob through quantum channel. 
                \end{enumerate}

{\bf Decryption: }\begin{enumerate}
                  \item Bob chooses measurement basis according to $k_1$:

                  while $k_1=0$, he takes measurement under $\{|0\rangle,|1\rangle\}$ basis, else if $k_1=1$, he takes measurement under $\{|+\rangle,|-\rangle\}$ basis. The result turns out to be $x\oplus k_2$ with probability {\rm 1}.
                  \item Bob gets $x=(x\oplus k_2)\oplus k_2$.
                \end{enumerate}
\end{protocol}
\subsection{{\bf Kind} 16}
The Kind 16 protocol requests: $P\in\mathbb{C},~~C,K,E,D\in\mathbb{Q}$.
This kind of protocol often uses EPR pairs $\frac{1}{\sqrt{2}}\left(|0\rangle_A|0\rangle_B+|1\rangle_A|1\rangle_B\right)$ as the key, each communication party shares one particle of the EPR pair.
According to the quantum nature of the key, it can not be copied at all.

We give an example protocol which encrypt one classical bit each time as follow:
\begin{protocol}[$P\in\mathbb{C},~~C,K,E,D\in\mathbb{Q}$]\label{protocol16}

~~~~~~~~~~~~~~~~~~~~~~~~~~~~~~~~~~~~~~~~~~~~~~~~~~~~~~~~~~~~~~~~~

Let key be $\frac{1}{\sqrt{2}}\left(|0\rangle_A|0\rangle_B+|1\rangle_A|1\rangle_B\right)$, and classical plaintext be $x$.

{\bf Encryption: }\begin{enumerate}
                  \item Alice prepares the quantum state $|\phi_x\rangle=|x\rangle_0$ according to $x$.
                  \item Alice performs CNOT operation on $|\phi_x\rangle$ with the particle in register A.
                  {\em\begin{eqnarray}
                    \frac{|0\rangle_A|0\rangle_B+|1\rangle_A|1\rangle_B}{\sqrt{2}}|x\rangle_0
                    \xlongrightarrow{\textrm{CNOT}}
                    \frac{|0\rangle_A|0\rangle_B|x\rangle_0+|1\rangle_A|1\rangle_B|x\oplus1\rangle_0}{\sqrt{2}}.
                  \end{eqnarray}}
                  \item Alice keeps the particle in register A, and sends the rest to Bob through quantum channel. 
                \end{enumerate}

{\bf Decryption: }\begin{enumerate}
                  \item Bob performs CNOT operation with the particle in register B on the qubit he received. It turns out
                  {\em\begin{eqnarray}
                    \frac{|0\rangle_A|0\rangle_B|x\rangle_0+|1\rangle_A|1\rangle_B|x\oplus1\rangle_0}{\sqrt{2}}
                    \xlongrightarrow{\textrm{CNOT}}
                    \frac{|0\rangle_A|0\rangle_B+|1\rangle_A|1\rangle_B}{\sqrt{2}}|x\rangle_0.
                  \end{eqnarray}}
                   \item Bob takes measurement to the qubit with $\{|0\rangle,|1\rangle\}$ basis, and gets $x$ with probability {\rm 1}.
                \end{enumerate}
\end{protocol}
\subsection{{\bf Kind} 28}
The Kind 28 protocol requests: $K\in\mathbb{C},~~P,C,E,D\in\mathbb{Q}$.
This kind of protocol aims at the quantum message, the most famous PQC(Private Quantum Channel) with unconditional security belongs to it. We give the process of PQC as follow:
\begin{protocol}[$K\in\mathbb{C},~~P,C,E,D\in\mathbb{Q}$]

For one bit quantum plaintext $\rho$, Alice and Bob need two bits classical key $k_1,k_2$.

{\bf Encryption: }\begin{enumerate}
                  \item Alice performs quantum operation $U_k=Z^{k_1}X^{k_2}$ to encrypt $\rho$ and gets $\rho'=U_k\rho U_k^{\dag}$.
                  \item Alice sends $\rho'$ to Bob through quantum channel.
                \end{enumerate}

{\bf Decryption: }\begin{enumerate}
                  \item Bob performs same quantum operation $U_k^{\dag}=Z^{k_1}X^{k_2}$ on $\rho'$ and get $\rho=U_k\rho' U_k^{\dag}$.
                \end{enumerate}
\end{protocol}
This protocol is proved unconditionally secure\cite{Ambainis00}.

\subsection{{\bf Kind} 32}
The Kind 32 protocol requests all five elements belong to quantum space. We show a simple example as follow:
\begin{protocol}[$P,C,K,E,D\in\mathbb{Q}$]

Let plaintext be $\rho$, and key be $\rho_k=\frac{1}{2}\left(|0\rangle_A|0\rangle_B+|1\rangle_A|1\rangle_B\right)\left(\langle0|_A\langle0|_B+\langle1|_A\langle1|_B\right)$. Alice obtains the particle in register A, and Bob obtains the particle in register B.

{\bf Encryption: }\begin{enumerate}
                  \item Alice performs CNOT operation on $\rho$ with her key:
                  \begin{eqnarray}
                    U_E(\rho_k\otimes\rho)U_E^{\dag}=\frac{1}{2}\sum_{a,b\in{0,1}}|a\rangle_A|a\rangle_B\langle b|_A\langle b|_B\otimes X^a\rho X^b.
                  \end{eqnarray}
                  \item Alice keeps the key state and sends the ciphertext state to Bob.
                \end{enumerate}

{\bf Decryption: }\begin{enumerate}
                  \item Bob performs CNOT operation on ciphertext state with his key. It turns out:
                  \begin{eqnarray}
                    U_D(\frac{1}{2}\sum_{a,b\in{0,1}}|a\rangle_A|a\rangle_B\langle b|_A\langle b|_B\otimes X^a\rho X^b)U_D^{\dag}=\rho_k\otimes\rho.
                  \end{eqnarray}
                  \item The plaintext is in the second register.
                \end{enumerate}
\end{protocol}
\subsection{Quantum key}
It should be mentioned that within the examples of Kind 16 and 32 protocol, the quantum key is entangled state rather than two identical but independent quantum states.
It makes a big difference from classical key.

In classical case, two parties only need to share two classical bit strings with same content, then the encryption and decryption operations run successfully. On the other hand, if two
parties share two identical but independent quantum states, the operations do not always work correctly. For example, assuming that Alice and Bob hold quantum state $\rho=|+\rangle$ as the key,
Alice encrypts $m=0$ according to the {\bf PROTOCOL \ref{protocol16}}, she will get:
\begin{eqnarray}
                    \frac{|0\rangle_A+|1\rangle_A}{\sqrt{2}}|0\rangle
                    \xlongrightarrow{\textrm{CNOT}}
                    \frac{|0\rangle_A|0\rangle+|1\rangle_A|1\rangle}{\sqrt{2}}.
\end{eqnarray}
She holds the quantum state in register A, and sends the ciphertext to Bob. Then Bob preforms decryption operation on it and gets:
\begin{eqnarray}
                    &&\frac{|0\rangle_B+|1\rangle_B}{\sqrt{2}}\frac{|0\rangle_A|0\rangle+|1\rangle_A|1\rangle}{\sqrt{2}}\nonumber\\
                    &\xlongrightarrow{\textrm{CNOT}}&\frac{|0\rangle_B(|0\rangle_A|0\rangle+|1\rangle_A|1\rangle)+|1\rangle_B(|0\rangle_A|1\rangle+|1\rangle_A|0\rangle)}{2}
\end{eqnarray}
The reduced density matrix of this result is
\begin{eqnarray}
\rho_m'=\frac{|0\rangle\langle0|+|1\rangle\langle1|}{2}=I,
\end{eqnarray}
which is a ultimate mixed state and has nothing to do with plaintext $m=0$.

The situation is assorted by that the encryption process with the classical key is uniform, so the decryption process with same key can decrypt successfully. while using the quantum key to preform
the encryption operation on $m$, if the key is a superposition state $|\phi_k\rangle=\sum_ip_i|k_i\rangle$ where each $|k_i\rangle$ controls a encryption process,
the result should be a mixture of ciphertexts: $\rho_c=\sum_i|p_i|^2E_i(|m\rangle)$, here $E_i(|m\rangle\langle m|)$ is the ciphertext encrypted by $|k_i\rangle$.
Then if Bob preforms decryption operation with same key $|\phi_k\rangle$, the final result is the mixture of all $D_j(E_i(|m\rangle\langle m|))$.
When $i=j$, $D_j(E_i(|m\rangle\langle m|))=|m\rangle\langle m|$, else $D_j(E_i(|m\rangle\langle m|))$ can collapse to any value.

On the other hand, if we choose entangled state as the key, the quantum encryption and decryption become feasible again. Since two parties share the entangle state, even if the encryption process is
a mixture of multiple processes, the decryption will turn back to the same state. In the meantime, the mixed encryption will strengthen the security of protocol.

It was Leung who presented the one time pad for quantum message with quantum key\cite{leung2000quantum}. In this protocol, the quantum key can be recycling used and security is ensured. This means that
even one qubit key can encrypt one qubit plaintext at once, the recycling property made the utilization efficiency of key high enough. Oppenheim
and Horodecki then proved that the classical key in PQC protocol can be used repeatedly and partly and it can also ensure security of the protocol as long as the detection is performed after each communication\cite{oppenheim2005reuse} .
\section{Type $\mathbb{O}$}
There are six kinds of protocols belong to type $\mathbb{O}$, which are not presented yet. We will give the structure ideas for two of them, and give the existence proof for the rest four of them.

\subsection{{\bf Kind} 2}
The Kind 2 protocol requests: $P,C,K,E\in\mathbb{C},~~D\in\mathbb{Q}$. The idea of construction is based on the classical NP problems. As there exists quantum resolution algorithms for some classical NP problems, we can use classical NP problem to do encryption, and use quantum resolution algorithm as decryption algorithm.

We take discrete logarithm problem for example, assume the resolution algorithm is $\mathcal{E}:\mathcal{E}(|a\rangle|b\rangle)=s$ for $b=a^s(mod~~q)$, and present the protocol as follow:
\begin{protocol}[$P,C,K,E\in\mathbb{C},~~D\in\mathbb{Q}$]

{\bf Encryption: }

Let plaintext be $x$, and key be $(g, h)$.
\begin{enumerate}
  \item Alice randomly chooses $y\in Z_q$, and calculates the ciphertext $(g^y,h^y\cdot m)$.
  \item Alice sends the ciphertext $(g^y,h^y\cdot m)$ to Bob through classical channel.
\end{enumerate}

{\bf Decryption: }\begin{enumerate}
                  \item Bob calculates $\mathcal{E}(|g\rangle|g^y\rangle)=y$.
                  \item Bob gets $m=h^y\cdot m/h^y$.
                \end{enumerate}
\end{protocol}

\subsection{{\bf Kind} 3}
The Kind 3 protocol requests: $P,C,K,D\in\mathbb{C},~~E\in\mathbb{Q}$. The construction idea is based on the
nature of quantum mechanics. We replace the classical random number algorithms with random collapse of quantum
measurement.

Assume $(P,C,K,\Lambda,E,D)$ is a classical probabilistic symmetric-key encryption
protocol\cite{yang2013quantum}. For example, according to the bit $x$, we randomly
choose $\lambda_x=b_1\cdots b_t$ satisfying $x=b_1\oplus\cdots\oplus b_t$, then
encrypt $\lambda_x$ and the ciphertext turns out to be $E(\lambda_x, k)$. This protocol
can hide the information of plaintext more effectively since each encryption process with the same plaintext turns out
different ciphertexts. Based on this protocol, we give an example
of the Kind 3 protocol as follow:
\begin{protocol}[$P,C,K,D\in\mathbb{C},~~E\in\mathbb{Q}$]

Let plaintext $x$ be a classical bit, and key be $k$.

{\bf Encryption: }
\begin{enumerate}
  \item Alice prepares quantum state $\sum_{\lambda_x}\frac{1}{\sqrt{2^{t-1}}}|\lambda_x\rangle$ based on $x$.
  \item Alice performs quantum encryption operation on $\sum_{\lambda_x}\frac{1}{\sqrt{2^{t-1}}}|\lambda_x\rangle$:
      $$U_E(\sum_{\lambda_x}\frac{1}{\sqrt{2^{t-1}}}|\lambda_x\rangle|k\rangle|0\rangle)=
      \sum_{\lambda_x}\frac{1}{\sqrt{2^{t-1}}}|\lambda_x\rangle|k\rangle|E(\lambda_x,k)\rangle.$$
  \item Alice measures the third register, the quantum state randomly collapses to $E(\lambda_x,k)$。
  \item Alice sends $E(\lambda_x,k)$ to Bob.
\end{enumerate}

{\bf Decryption: }
\begin{enumerate}
    \item Bob calculates $D(E(\lambda_x,k),k)=\lambda_x$ with key $k$ and gets $x$ which is the parity bit of $\lambda_x$.
\end{enumerate}
\end{protocol}

\subsection{Rest ones}
For the rest four kinds of type $\mathbb{O}$, Kind 4, 8, 20, 24, we give the existence proofs for them.
\begin{enumerate}
\item The Kind 4 protocol requests $P,C,K\in\mathbb{C},~~E,D\in\mathbb{Q}$. We can improve a simple classical symmetric-key encryption into this kind as follow:

Take quantum circuits which realize the equivalent computational process to the classical encryption and decryption algorithms as the quantum encryption and decryption ones.
Since $P,C,K\in\mathbb{C}$, the input plaintext and key for quantum encryption algorithm will firstly be coded with $|0\rangle,|1\rangle$ basis, and the output ciphertext will be measured with
$|0\rangle,|1\rangle$ basis too. For quantum decryption algorithm, the same process will be performed. Hence it comes out a Kind 4 protocol.

\item The Kind 8 protocol requests $P,C\in\mathbb{C},~~K,E,D\in\mathbb{Q}$. It comes out to be the same as the Kind 4 protocol except we take the quantum state coding from the
classical key as the new protocol's key.

\item The Kind 20 and 24 protocols can be constructed as the same.
 \end{enumerate}

However, this modification scheme can not take advantage on the security or practicability, it only proves the existence of these kinds of protocols.
We hope some meaningful protocols can be presented in the future.

\section{Type $\mathbb{N}$}
As we shown in Tab.1, there are 21 kinds of quantum symmetric-key encryption considered to be unable to construct, including the Kind 5-7, 9-11, 13-15, 17-19, 21-23, 25-27 and 29-31 protocols.

These kinds of protocol have one trait in common: they request at least one of the encryption and decryption algorithms belongs to classical space, meanwhile they request at least one of the
plaintext, ciphertext, and key belongs to quantum space. This means they will take a quantum state as the input or output object of a classical algorithm, which is not valid.

If a algorithm involve a quantum object, it must also belong to quantum space. Both encryption and decryption algorithms involve plaintext, ciphertext and key simultaneously.
Encryption algorithm takes plaintext and key as input objects, and ciphertext as output object. Decryption algorithm takes ciphertext and key as input objects, and plaintext as output object.
As a result, if one of the encryption and decryption algorithms belongs to classical space, the plaintext, ciphertext and key must all belong to classical space.
Therefore these 21 kinds of quantum symmetric-key encryption are unable to construct.
\section{Discussion}
In this article there are six kinds of protocols belong to type $\mathbb{O}$. These protocols are only given theoretical structure scheme, which are inefficiency
and not practical. How to find an efficient structural scheme or prove there are no effective ones is in our further study. In addition, we also study the classification of
quantum public-key encryption protocol. Since its key has two parts to be considered, it must be classify under six-tuple $(P,C,G_1,G_2,E,D)$. For example, Min Liang and Li Yang present a
quantum public-key encryption protocol\cite{liang2012public,liang2012quantum}, the secret key is a classical function $F$, the public key is a quantum state $\rho_{k_j}$ and a classical bit string $s_j$.

\section{Conclusions}
Quantum symmetric-key encryption protocol is generalized concept and includes the classical one in this article. Based on the quintuple of it, we classify the 32 kinds of quantum symmetric-key encryption
protocols into three types. First type includes five kinds of protocols that all have been presented yet and worth deeply studying. Second type includes six kinds of protocols that are only proved exist
theoretically and worth discussing. The last type includes the rest 21 kinds, they are proved to be unable to construct as they request classical algorithm work on quantum target, which is not valid. We also
suggest that the quantum key has a big difference from the classical key. Two parties should share two parts of an entangled state rather than two identical but independent quantum states, otherwise the
encryption and decryption operations do not always work correctly.

\section*{Acknowledgement}
This work was supported by the National Natural Science Foundation of China under Grant No.61173157.

\end{document}